\documentstyle{article}
\begin{document}
\begin{center}
{New type of extreme value statistics}\\
{D.B. Saakian}\\
{Yerevan Physics Institute,
Alikhanian Brothers St. 2, Yerevan 375036, Armenia.}

\end{center}

\begin{abstract}
We investigate the extreme value statistics connected with the 
dilute Random Energy Model with integer couplings. 
New universality class is found.
\end{abstract}
\vspace{10mm}
Extreme value statistics [1]  statistics play an important role in 
statistical
mechanics and theory of probability [2-5]. They describe the
minimal and maximal bounds 
for the distribution of a large number
of independent random quantities, and it is the main reason
of their recent applications in statistical physics of 
disordered systems.\\
Usually one considers sequence of $M=2^N$ independent  random  variables 
$E_i$ with the same distribution
\begin{equation}
\label{e1}
\rho(E)=<\delta(E_i-E)>
\end{equation}
Let us find minimal and maximal members $E_{min}$ and $E_{max}$ 
of our sequence,
then consider their distribution at the limit of large N.
When $\rho(E)$ is independent from N, at $N\to \infty$ limit only 
3 classes of universalities has been found  
[1] for the distribution of $\rho(E_{min})$.
After rescaling of $x\equiv E_{min}$ one has for them:
\begin{equation}
\label{e2}
P_1(x)=1,x\ge 0, P_1(x)=1-\exp[-(-x)^k)]\qquad  x\le0
\end{equation}
\begin{equation}
\label{e3}
P_2(x)=0,x\le 0, P_2(x)=1-\exp[-x^k)]\qquad  x\ge0
\end{equation}
\begin{equation}
\label{e4}
P_3(x)=1-\exp[-e^x]
\end{equation}
Now we are going to consider another (physically interesting) 
case, when our $\rho(E)$ is a superposition of 
N random variables. \\
In the present paper we will investigate extreme value statistics which are
generated by the dilute-coupling REM. Surprisingly, we got new types
of them due to physical peculiarities of the model.\\
The dilute coupling REM  is described by the hamiltonian 
\begin{equation}
\label{e5}
H=-\sum_{(i_1..i_p)=1}^{\alpha N}j_{i_1..i_p}s_{i_1}..s_{i_p}
\end{equation}
where only $\alpha N$ quenched bonds are nonzero,
 and they take values $\pm  1$
with equal probability. $s_i=\pm 1$ are the Ising spins.\\
In the limit of large p there is a factorization in probability of different energy levels and we have  $2^N$
independent random variables (energies)  with the same probability distribution
$\rho(E)$.  
For our choice of integer couplings we have for
distribution $\rho(E)$.
\begin{equation}
\label{e6}
\rho(E)=\frac{1}{2\pi i }\int_{-i\infty}^{i\infty}d ke^{N\varphi(k)-kE}
\end{equation}
where
\begin{equation}
\label{e7}
\varphi(k)=\alpha\ln \cosh(k)
\end{equation}
This representation is correct for
\begin{equation}
\label{e8}
|E|<\alpha 
\end{equation}
 Otherwise $\rho(E)$ is strictly zero. \\
For the case $\alpha<1$ we should keep in mind, that there are $2^{\alpha N}$  different classes of energy 
configurations with  
\begin{equation}
\label{e9}
2^{(1-\alpha)N} 
\end{equation}
equal energies in each one.  Energies from different classes are distributed independently 
according to (6).\\
At low temperatures the system is frozen in minimal energy configuration (chozen among the all
 $2^N$ ones).
Thats why the physics of REM is connected closely  with extreme value statistics problem. \\
As known, several phases are possible in the considered model.
For  $\alpha >1$
there is a phase transition to the spin glass phase and at high 
temperatures 
the system is in the
paramagnetic phase. For the case $\alpha <1$
only the paramagnetic phase is possible. The limiting case
$\alpha \sim 1$
corresponds to the border between 2 phases.\\
We hope, that different universality classes are possible, connected with those phases.
Let us consider $2^N$ independent random variables with  (6), where  $\varphi(k)$ is  even analytical function 
with respect to $k$.
It is just enough for the most of situations.
For the case of hamiltonian (5) one has, that 
\begin{equation}
\label{e10}
\lim _{k\to \infty}\varphi(k)=\alpha (k-\ln2)+o(\exp(-2k))
\end{equation}
We are considering only distributions , which have this asymptotic form,
or when corrections are polynomial
\begin{equation}
\label{e11}
\lim _{k\to \infty}\varphi(k)=\alpha(k-\ln2)+o(k^{-\beta})
\end{equation}
We are going to consider now the  extreme value statistics for $2^N$ random variables with our representation
for $\rho(k)$.\\
Let us define the probability $\theta(E)$, as the integral of these
expressions in the interval $[-\infty,E]$. We have 
\begin{equation}
\label{e12}
\theta(E)=\frac{1}{2\pi i }\int_{-i\infty}^{i\infty}d
ke^{N\varphi(k)-kE}/k
\end{equation}
Here our integration loop overpasses the point 0 from the left size.\\
Let us find the distribution of the minimal bound. We take one variable $E_i$ 
in our
interval $[-\infty,E]$, the others to right. As the first one could be
chosen from the all $2^N$, we have an expression [4]
\begin{equation}
\label{e13}
P(E)=2^N \rho(E)(1-\theta(E))^{2^N-1}=-\frac{d}{dE}(1-\theta(E))^{2^N}
\end{equation}
We need to study the expression $\theta(E)$  in the region, when it is small. Here 
\begin{equation}
\label{e14}
(1-\theta(E))^{2^N}\approx \exp[ -2^N\theta(E)]
\end{equation}
Its density  is nonzero only near the $E=E_c$. We need to find
asymptotic of function $\theta(E)$ near the $E_c$, defined by condition
\begin{equation}
\label{e15}
2^N\theta(E_c)= 1
\end{equation}
For the different choices of $\alpha$ we have to consider different asymptotic regims for $\theta(E)$.
In this way different universality classes arose. \\
 While calculating (12) for the saddle point we have an equation
\begin{equation}
\label{e16}
\varphi'(k)=-\frac{E}{N}
\end{equation}
There are three possibilities for $E_c$.
The first one 
\begin{equation}
\label{e17}
|E_c|<\alpha N
\end{equation}
has the same statistics, as in the case when  $\varphi(k)$ 
 goes to infinity faster  than 
linearly   (this is the case of ordinary REM [4] with normal
distribution of energies). Two others are possible also:
\begin{equation}
\label{e18}
|E_c|\sim\alpha N
\end{equation}
\begin{equation}
\label{e19}
|E_c|>\alpha N
\end{equation}
In the last case we should take complex solutions of (16).
Let us first calculate the first case (18). Two expressions for
the function $\theta(E)$ coincide (at  the limit of large N) and we have 
\begin{equation}
\label{e20}
\theta(E)=\frac{1}{k\sqrt{ 2 \pi N \varphi''(k)}}e^{N\varphi(k)-kE}
\end{equation}
One can write the expression for the desired probability distribution for
the minimum of our $2^N$ random variables:
\begin{equation}
\label{e21}
e^{\left (
-\frac{2^N}{k \sqrt{ 2 \pi N \varphi ''(k)}} 
e^{N\varphi(k)-kE}
\right )}
\frac{d}{dE}\frac{1}{k  \sqrt{ 2 \pi N \varphi '' (k)  }     }
e^{N\varphi(k)-kE}
\end{equation}
For our asymptotic form (15) transforms into
\begin{equation}
\label{e22}
\frac{2^N}{k\sqrt{ 2 \pi N \varphi''(k)}}e^{N\varphi(k)-kE_c}=0
\end{equation}
where  k is defined from (16).
For the $E<E_c$ our probability is almost one, and above it is almost zero.
For a small deviation 
\begin{equation}
\label{e23}
u\equiv(E-E_c)k_c\ll N
\end{equation}
we obtain Gumbel distribution
\begin{equation}
\label{e24}
P(u)=\exp(u-\exp(u)),
\end{equation}
where $u$ changes in the interval $-\infty,\infty$.
Let us now consider the statistics for limiting situation, when $\alpha\sim 1$.
 We shall consider the distribution (7), but 
the corresponding limiting
statistics is the same for the whole class of exponential corrections.
At the saddle point 
\begin{equation}
\label{e25}
\tanh(k)=E/(\alpha N )
\end{equation}
we have 
\begin{equation}
\label{e26}
\theta(E)=\frac{1}{2\pi i }\int_{-i\infty}^{i\infty}e^{\alpha N\ln\cosh(k)-kN}/k\approx 
\end{equation}
$$=\frac{\cosh(k)}{k\sqrt{N2\pi}}e^{-\alpha N g(E/(\alpha N)}$$
where  
\begin{equation}
\label{e27}
g(x)=\frac{1+x}{2} \ln \frac{1+x}{2} +\frac{1-x}{2} \ln \frac{1-x}{2} 
\end{equation}
Let us do notation
\begin{equation}
\label{e28}
N_1=N\alpha,\frac{E}{\alpha N}=1-\epsilon
\end{equation}
We are considering the case, where $\frac{N_1}{N}-1\equiv\epsilon_1\to 0$.
In the limit $\epsilon\to 0$ we have 
\begin{equation}
\label{e29}
g(1-\epsilon)\to\ln 2-\frac{ \epsilon}{2} \ln (\frac{2}{\epsilon }),\frac{1}{\cosh(k)^2}\sim \epsilon
\end{equation}
\begin{equation}
\label{e30}
\theta(E)=
\frac{1}{\ln (  \frac{2}{ \epsilon }  )    \sqrt{   N\pi \epsilon    } }
\exp \left ( - N _1\ln 2 + \frac{N_1\epsilon}{2} \ln (2 /  \epsilon ) \right  )
\end{equation}
Let us check, when such a behaviour is correct. 
We should take condition
\begin{equation}
\label{e31}
\frac{N}{\cosh^2(k)} \sim N  \epsilon  \gg 1
\end{equation}
Now let us consider the region $N\epsilon \sim 1$.
Of course,  at the saddle point we cannot approximate our integral by the
gaussian one for our values of $\epsilon$,  since  the second derivative 
tends to zero. We should directly calculate integral.\\
  For the complex  
$k\equiv k_1
+ik_2$ we have
\begin{eqnarray}
\label{e32}
&&\ln\cosh(k)=\cosh k_1+1/2\ln [1-(1-\tanh^2k_1)\sin^2k_2]
\nonumber \\
&&+i\arctan[\tan k_2-
(1-\tanh k_1)\tan k_2]
\end{eqnarray}
This is an exact expression. Let us consider the case $k_1\to \infty$.
In this limit
\begin{equation}
\label{e33}
\ln\cosh(k)=k_1-\ln2+e^{-2k_1},\ \  \tanh k_1=1-2e^{-2k_1}
\end{equation}
So we have to choose the position of $k_1$, while performing integration 
\begin{equation}
\label{e34}
\frac{1}{2\pi i
}\int_{k_1-i\infty}^{k_1+i\infty}e^{\alpha N[\ln\cosh(k)-(1-\epsilon]k}/k
\end{equation}
The saddle point equation gives the choice
\begin{equation}
\label{e35}
k_1=\frac{1}{2}  \ln (  \frac{2}{\epsilon}  )
\end{equation}
With such a choice of $k_1$ we get
\begin{equation}
\label{e36}
N_1\ln\cosh(k)-N_1(k_1+ik_2)+N_1\epsilon(k_1+ik_2)
\end{equation}
$$\approx N(\epsilon/2+\epsilon/2\ln 2N/\epsilon-\epsilon\sin 
k_2+i\epsilon(k_2-\sin k_2\cos k_2))$$
Accuracy of this expression is
$\exp (-2k_1)\sim \epsilon $.
Our expressions are valid in  a region
\begin{equation}
\label{e37}
N\gg \epsilon
\end{equation}
We get
\begin{eqnarray}
\label{e38}
&&\theta(E)\approx
\exp \left ( -N_1\ln 2 +  \frac{ N\epsilon}{2}  \ln \left  (  \frac{ 
2}{\epsilon}\right  )\right )\nonumber \\
&&\frac{1}{\pi}\int_{o}^{\infty}d k_2 (k_1^2+k_2^2)^{-1}e^{-\epsilon\sin 
k_2^2}
k_1\cos(\epsilon(k_2
\nonumber \\
&&-\sin k_2\cos k_2))+k_2\sin(\epsilon(k_2-\sin
k_2\cos k_2))         
\end{eqnarray}
For the limit of small $\epsilon$ last expression transforms into
\begin{equation}
\label{e39}
\theta(E)=
\exp{(-N_1\ln 2+\frac{N\epsilon}{2} \ln ( 2/\epsilon))}
\end{equation}
We have for the desired distribution for $u=
(\frac{E}{\alpha N}-1)N$
\begin{equation}
\label{e40}
P(u)=\left ( -1+\ln\frac{ N}{u} \right )\exp \left [ u\ln\frac{ 
N}{u}-\exp \left (u\ln\frac{ N}{u} \right ) \right ]
\end{equation}
Such distribution is correct in the interval $-\infty<u<0$. What about
negative values of u? For the choice (7 )with integer $\alpha N$ we
have, 
that there is a strict bound for energy distribution equal to $N\alpha$,
so for that case we can put zero probability for negative values of u .
This expression should exist for any bounded discrete
distribution of energies.\\
In [5] authors consider fully connected version of REM, here perhaps could be only classic
distribution among the (2)-(4). The case of bounded continuous
distribution brings to (2)-(3).\\
This work was supported by Fundacion Andes grant c-13413/1. I would like
to thank  E. Vogel for hospitality in UFRO.

\end{document}